\documentclass[10pt]{iopart}  
\usepackage{graphicx}
\usepackage{epsfig}

\begin{document}

\def\be{\begin{equation}}
\def\ee{\end{equation}}
\def\ba{\begin{eqnarray}}
\def\ea{\end{eqnarray}}
\def\bas{\begin{eqnarray*}}
\def\eas{\end{eqnarray*}}
\def\pra{\PR A}
\def\prb{\PR B}
\def\prc{\PR C}
\def\prl{\PRL}

\title[DMRG and wave function factorization]
{Density matrix renormalization group and wave function factorization 
for nuclei}

\author{T. Papenbrock\dag\ddag\ and D. J. Dean\ddag }

\address{\dag\ Department of Physics and Astronomy, University of Tennessee,
Knoxville TN 37996-1201, USA}

\address{\ddag\ Physics Division,
Oak Ridge National Laboratory, Oak Ridge, TN 37831, USA}

\begin{abstract}
We employ the density matrix renormalization group (DMRG) and the wave
function factorization method for the numerical solution of large-scale 
nuclear structure problems. The DMRG exhibits an improved convergence for
problems with realistic interactions due to the implementation of the
finite algorithm. The wave function factorization of $fpg$-shell nuclei
yields rapidly converging approximations that are
at the present frontier for large-scale shell-model calculations.    
\end{abstract}

\pacs{21.60.Cs,21.10.Dr,27.50.+e}

\submitto{\JPG}


\section{Introduction}
The nuclear shell model with full configuration mixing is the adequate
tool for a quantitative description of light- and medium-mass
nuclei. In recent years, the no-core shell model has successfully been
applied to investigate $p$-shell nuclei \cite{Nav00a,Nav02}, and full-space
diagonalizations involving model spaces with up to one-billion Slater
determinants are now possible for $sd$-shell and $pf$-shell nuclei
within the traditional shell model \cite{Cau94,Cau99,Hon02}. 
Note that it took more than a
decade to go from $sd$-shell nuclei to $pf$-shell nuclei, and progress
was due to more sophisticated algorithms and an increase in
computer power. The description of heavier nuclei or drip-line nuclei,
however, is based on increasingly larger model spaces. For such
nuclei, exact diagonalizations will be unavailable for the next 
years, and one has to introduce approximations.

Several authors have suggested truncations of the model space. In many
of these methods, the selection of the relevant basis states is based
on physical insights and arguments and is done ``by hand''
\cite{Hor94,And01,Gue02}. In other methods, the Hamiltonian itself
selects the most important basis states. One example is the Monte
Carlo shell model \cite{Hon95}, where the huge Hilbert space is
sampled stochastically, and only relevant basis states are
kept. Another example is provided by the density matrix
renormalization group (DMRG) \cite{Whi92,Whi93} and the wave function
factorization \cite{Pap03}, where the most important basis states are
determined from a variational principle. It is the purpose of this
article to describe recent progress with the latter two methods. We
report on DMRG results for $sd$-shell and $pf$-shell nuclei, and use
the wave function factorization to compute accurate approximations for
low-lying states in $0f_{5/2}\,1p\,0g_{9/2}$ shell nuclei that are at
the frontier of full space diagonalizations.

\section{Density matrix renormalization group}
During the last decade, the DMRG has become the method of choice to
compute ground state properties for spin-chains. We refer the reader
to the review \cite{Pes99} and briefly describe the main aspects of this
method. In the one-dimensional spin systems, the lattice is divided
into two parts (e.g. ``left'' and ``right''), and the ground state
energy is computed for a small number of lattice sites.  The ground
state can be expressed in terms of basis states $|l\rangle$ and
$|r\rangle$ of the ``left'' and ``right'' part of the chain,
respectively, as
\be
\label{svd}
|\Psi\rangle=\sum_{l,r=1}^M\Psi_{l r}|l\rangle|r\rangle.
\ee
In the DMRG, one seeks an approximation of $|\Psi\rangle$ in terms of
$m<M$ basis states. These states are the eigenstates of the density
matrices $\rho_{l l'}=\sum_r \Psi_{l r}\Psi_{l' r}$ and $\rho_{r
r'}=\sum_l \Psi_{l r}\Psi_{l r'}$ that correspond to the $m$ largest
eigenvalues. This is a truncation for which the difference between
exact ground state and approximated ground state has minimal norm
\cite{Whi93}. One then retains these $m$ states for the left and right
part of the lattice, and adds another lattice site to either
part. This procedure defines the ``infinite algorithm'' and is
repeated until the desired system size is reached. At this point, an
iterative procedure known as the ``finite algorithm'' is used to
enlarge one part of the lattice at the expense of the other while
keeping the total number of lattice sites constant. This defines a
``sweep'' through the system. Typically one finds that the truncation
error decreases exponentially with the number $m$ of kept states. Note
that one does not store the wave function in the DMRG. Instead, one
keeps $m$-dimensional matrix representations of those operators that
define the Hamiltonian and other observables of interest.

Only recently has the DMRG also been applied to finite Fermi systems,
and we refer the reader to the recent review by Dukelsky and Pittel
\cite{Duk04}.  There are at least three novelties for the DMRG in
finite Fermi systems. First, an order of lattice sites,
i.e. single-particle orbitals has to be chosen. Second, a partition of
the system in two parts has to be chosen. Third, the two-body
interaction induces interactions between very distant single-particle
orbitals and thereby differs from the spin chains with neighboring
interactions. For nuclei, Dukelsky and coworkers suggest an ordering
of the single-particle orbitals based on their distance from the Fermi
surface and based the partition on particle versus hole orbitals
\cite{Duk02}. Using the infinite DMRG algorithm, they obtained rapidly
converging results for schematic pairing and pairing-plus-quadrupole
models but encountered rather slow convergence for $sd$-shell nuclei
with more realistic (and complex) interaction \cite{Dim02}.
 
Our implementation of the DMRG is as follows. We choose a partition of
neutron orbitals versus proton orbitals.  This is motivated by our
experience with the wave function factorization.  The spherical single
particle orbitals have quantum numbers $\alpha=(n,l,j,j_z,\tau_z)$. We
work in the $m$ scheme and conserve total angular momentum $J_z$ at
each step of the DMRG. As the equivalent of a single lattice site, we
choose conjugate pairs $(\alpha,\bar{\alpha})$ of single-particle
orbitals. This is designed to improve angular momentum properties and
to properly treat pairing correlations.  The conjugate pairs of
single-particle orbitals are ordered such that the most active
orbitals, i.e. the orbitals close to the Fermi surface are in the
center of the chain of orbitals. This is motivated by the extensive
studies reported by Legeza on this subject \cite{Leg03}.  To overcome
difficulties with the long-ranged two-body interaction, we use the
finite algorithm. We do not use the infinite algorithm. Instead, we
start the finite algorithm based on the spherical Hartree-Fock configuration
and its $1p$-$1h$ excitations.  We find that this approach is superior
to the infinite algorithm and the approach suggested by Xiang for
treating systems with conserved quantities \cite{Xia96}. Note that we also
considered DMRG calculations in a Hartree-Fock basis. Compared to the
DMRG in the spherical basis, this approach lowered the energy at the
start of the finite algorithm.  However, the breaking of axial
symmetry led to larger-dimensional matrix problems and did not improve
the convergence of the method. In the practical calculations, we start
with the finite algorithm with a rather small number $m$ of kept
states and increase this number every two sweeps.

\begin{figure}[hbpt]
\includegraphics[scale=0.23]{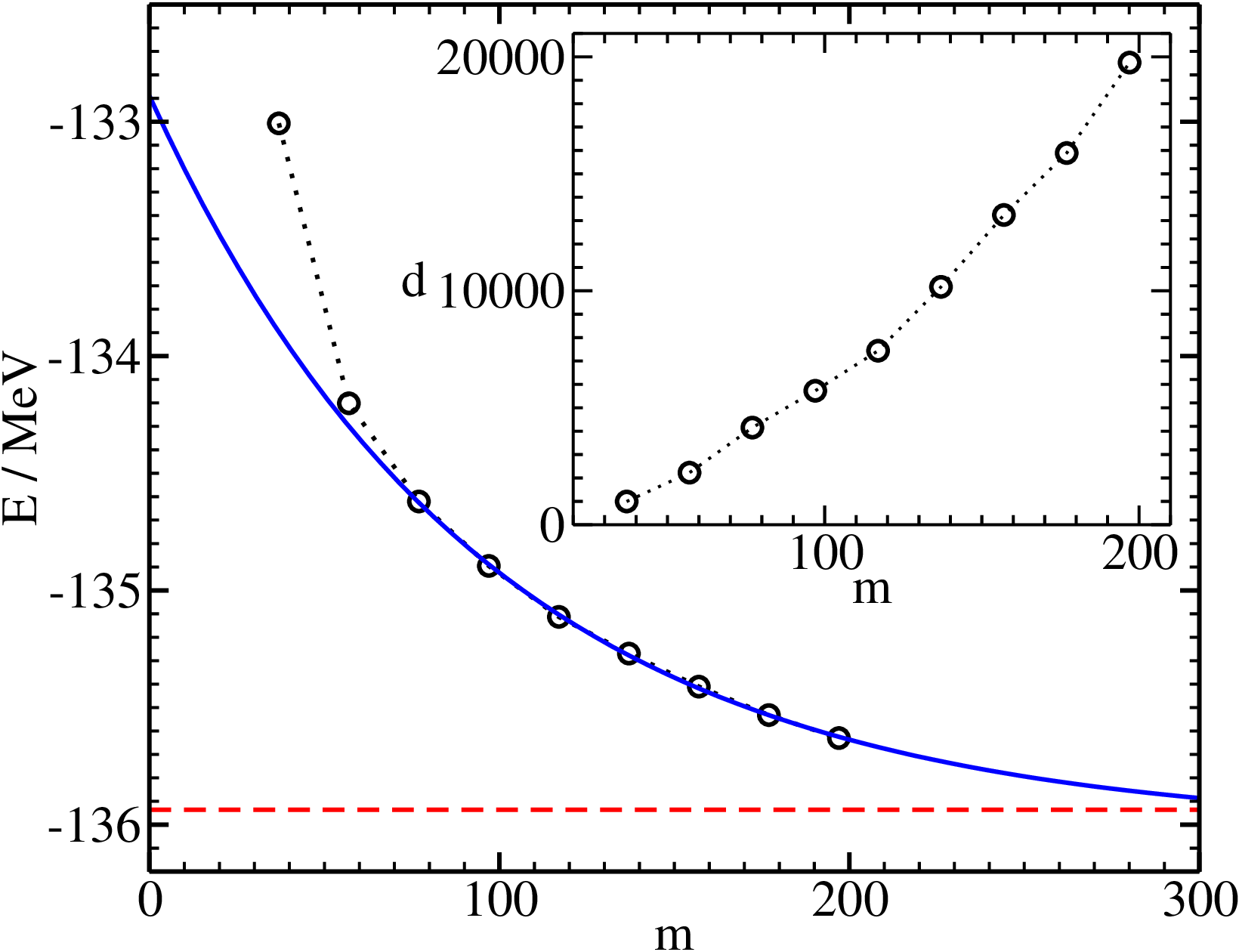}
\includegraphics[scale=0.23]{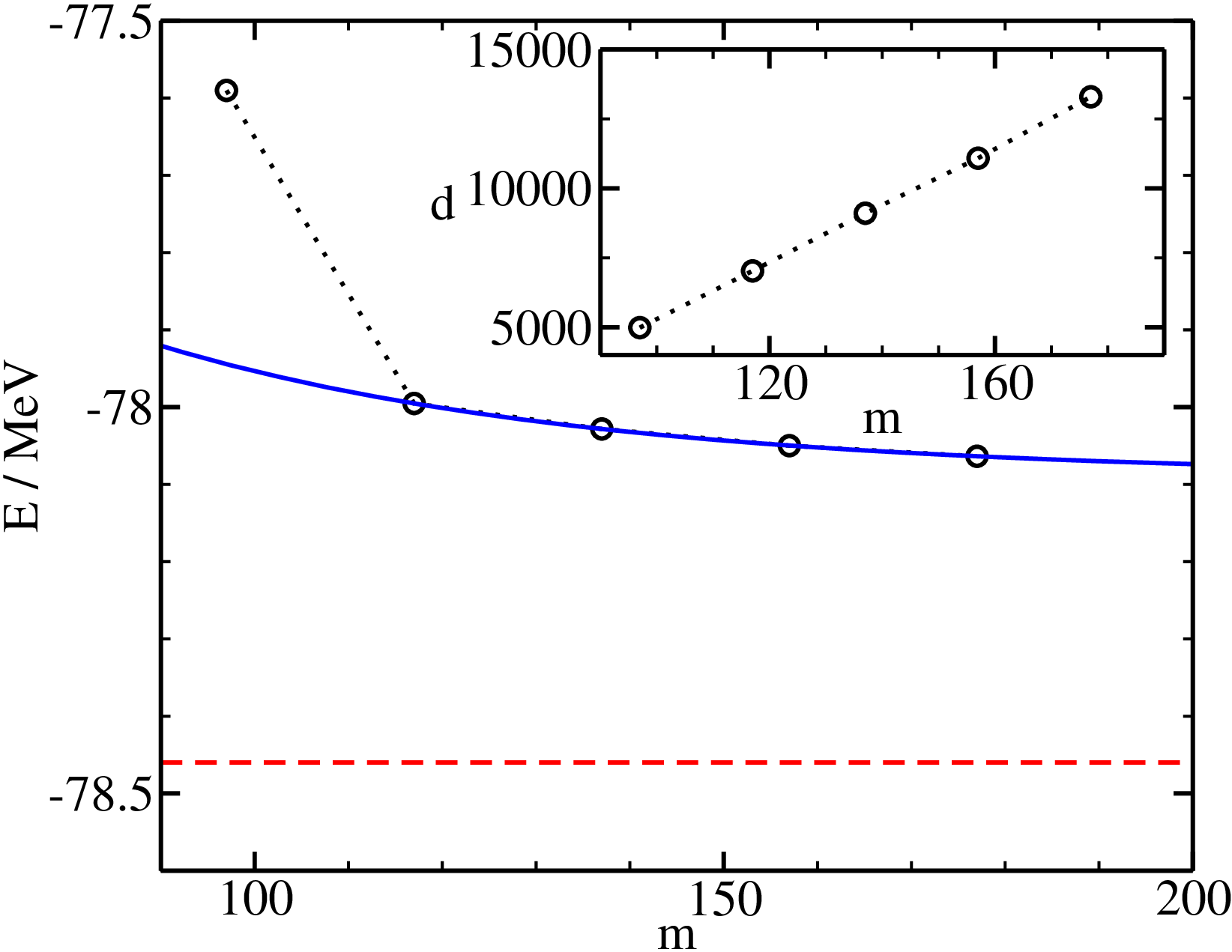}
\caption{\label{fig1}Left: DMRG ground state energies for $^{28}$Si as
a function of the number $m$ of kept states (data points connected by
dots), exact result (dashed line), and exponential fit (full line). 
The inset shows the dimension $d$ of the DMRG eigenvalue problem vs. $m$.
Right: Same as left except for $^{56}$Ni.}
\end{figure}

We performed DMRG calculations for the $sd$-shell nucleus $^{28}$Si
using the USD interaction \cite{Bro88}. The order of the (pairs) of
single-particle orbitals is $d_{3}3$, $d_{3}1$, $d_{5}5$, $d_{5}3$,
$s_{1}1$, $d_{5}1$, $d_{5}1$, $s_{1}1$, $d_{5}3$, $d_{5}5$, $d_{3}1$,
$d_{3}3$, where we used the notation $l_{2j} |2j_z|$, and it is
understood that the first half of orbitals corresponds to proton
states, while the (mirror symmetric) second half of orbitals
corresponds to the neutron states. The initial set of states consists
of the $(d_{5/2})^{12}$ configuration and its $1p$-$1h$
excitations. The left part of Fig.~\ref{fig1} shows the DMRG results
for the ground state energy as a function of the number of states we
keep.  The energy converges exponentially rapidly as more states are
kept, and the true ground state energy can accurately be determined
from an exponential fit to the data. The inset shows the dimension of
the DMRG eigenvalue problem as a function of the number of kept
states. Note that the full eigenvalue problem has dimension $d=93710$.

This promising result motivates us to perform DMRG calculations for
the $fp$-shell nucleus $^{56}$Ni. We use the KB3 interaction
\cite{Kuo68,Pov80}. The exact diagonalization of this shell-model
problem has to deal with about one billion Slater determinants and has
only been accomplished recently \cite{Cau99}. 
The order of single-particle
orbitals is $f_{5}5$, $f_{5}3$, $f_{5}1$, $p_{1}1$, $f_{7}7$,
$f_{7}5$, $p_{3}3$, $f_{7}3$, $p_{3}1$, $f_{7}1$, $f_{7}1$, $p_{3}1$,
$f_{7}3$, $p_{3}3$, $f_{7}5$, $f_{7}7$, $p_{1}1$, $f_{5}1$, $f_{5}3$,
$f_{5}5$. The initial set of states consists of the $(f_{7/2})^{16}$
configuration and its $1p$-$1h$ excitations. The right part of
Fig.~\ref{fig1} shows the results. The DMRG results seem to be
converged as an increase of $m$ does not significantly lower the
energy. However, the true ground state energy is almost 400 keV lower
in energy. We recall that the structure of $^{56}$Ni is to some extent
similar to $^{28}$Si. This similarity has motivated our choice of
states to be kept at the beginning of the finite algorithm and our
order of the single-particle orbitals. It is thus unexpected that the
DMRG converges for $^{28}$Si but fails to fully converge for the
computationally larger problem $^{56}$Ni.

Note however that our DMRG computations exhibit an improved
convergence when compared to the pioneering study \cite{Dim02}. The
most important new ingredients are the use of the finite algorithm,
the abandoning of the infinite algorithm, and the ordering of the
single-particle orbitals. We expect that further improvement of the
convergence relies on an optimal order of single-particle orbitals and
on optimal initial conditions for the finite algorithm. The method
presented in the following section avoids these issues and makes use
only of the unproblematic truncation step of the DMRG.

\section{Wave function factorization}

Modern shell model codes build their basis from products of Slater
determinants $|\pi\rangle$ for the proton space and Slater determinants
$|\nu\rangle$ for the neutron space. Note that the dimensions of the
proton and neutron space are modest compared to the dimension of the
product space. A shell-model ground state might thus be expanded as
\be
\label{pnsvd}
|\Psi\rangle=\sum_{\pi,\nu}\Psi_{\pi \nu}|\pi\rangle|\nu\rangle,
\ee
where the sum runs over all available proton and neutron determinants.
This expansion is similar to Eq.~(\ref{svd}) used in the DMRG.
The amplitude matrix $\Psi_{\pi \nu}$ is rectangular in general and can be
factored as $\Psi=USV^\dagger$ by means of the singular value
decomposition (SVD).  Here, $U$ and $V$ are unitary matrices that operate
on the proton space and the neutron space, respectively, and $S$ is
zero except for nonnegative entries $s_{jj}\ge 0$ (the ``singular values'')
along its diagonal.
The SVD thus yields correlated proton and neutron states 
\ba
|\tilde{p}_j\rangle &=& \sum_\pi U_{\pi j} |\pi\rangle,\\
|\tilde{n}_j\rangle &=& \sum_\pi V_{\nu j} |\nu\rangle,
\ea
and the ground state becomes $|\Psi\rangle=\sum_j s_j |\tilde{p}_j\rangle
|\tilde{n}_j\rangle$.  Note that the DMRG is closely related to the SVD:
$\Psi\Psi^\dagger = USS^\dagger U^\dagger$ and $\Psi^\dagger\Psi =
VS^\dagger SV^\dagger$ are density matrices which are diagonalized by
the matrices $U$ and $V$, respectively, and their eigenvalues are the
squares of the singular values \cite{Whi93}. Note also that the SVD yields
exponentially rapidly decreasing singular values for ground states
obtained from realistic shell-model interactions \cite{Pap04}. 
This observation is interesting for two reasons. First, it shows that 
the DMRG truncation step also works for nuclei. Second, it provides
the basis for a powerful approximation and basis state selection scheme.
In the wave function factorization, we approximate
 the ground state as a sum over
$\Omega$ products of correlated proton and neutron states 
\be
\label{factor}
|\Psi\rangle= \sum_{j=1}^\Omega |p_j\rangle |n_j\rangle. 
\ee 
Note that the states $|p_j\rangle$  and $|n_j\rangle$ are not normalized. 
Variation of the energy yields the following set of eigenvalue equations 
that determine the states $|p_j\rangle$  and $|n_j\rangle$
\ba
\label{solution}
\sum_{i=1}^\Omega\left(\langle n_j|\hat{H}|n_i\rangle - E\langle
n_j|n_i\rangle\right)|p_i\rangle &=& 0,\nonumber\\
\sum_{i=1}^\Omega\left(\langle p_j|\hat{H}|p_i\rangle - E\langle
p_j|p_i\rangle\right)|n_i\rangle &=& 0.  
\ea 
These equations can be solved iteratively. One starts with a random
set of orthogonal neutron states $|n_j\rangle$ and solves the first of
the eigenvalue problems in Eq.~(\ref{solution}) for those proton
states that correspond to the lowest energy $E$. These are then
inserted in the second eigenvalue problem in Eq.~(\ref{solution}),
which is solved for the neutron states corresponding to the lowest
energy. This procedure is repeated until the energy $E$ converges. For
even-even nuclei, convergence is typically reached in 5-10 iterations,
while other nuclei require a somewhat larger number of iterations.
The eigenvalue problems~(\ref{solution}) have dimension $\Omega D_p$
and $\Omega D_n$ for the proton states and neutron states,
respectively, where $D_p$ ($D_n$) denotes the dimension of the proton
(neutron) space. Typically, one needs only a few hundred states
$\Omega$ for more than 99\% overlap with the exact ground state. Thus,
$\Omega \ll D_p, D_n$, and the factorization requires one to solve
eigenvalue problems of small dimensions relative to the exact
diagonalization which has a dimension that scales as $D_p D_n$.  Note
that the proton and neutron states that enter the eigenvalue
problem~(\ref{solution}) can be kept orthogonal, and this reduces the
generalized eigenvalue problem to a standard eigenvalue problem. Note
also that axial or rotational symmetry can be conserved within the
ansatz~(\ref{factor}). For details, we refer the reader to
Ref.~\cite{Pap04}.

In earlier work, we applied the factorization to $sd$-shell nuclei and
$fp$-shell nuclei and demonstrated its accuracy and capability by
comparing with results from exact diagonalization.  In this work we
use the factorization and perform structure calculations for a few
$A=76,78$ nuclei. The model space consists of the
$0f_{5/2}\,1p\,0g_{9/2}$ orbitals, and the interaction is a
monopole-corrected $G$-matrix \cite{Shu83}, taking
$^{56}_{28}$Ni$_{28}$ as a closed core.  These problems are at the
present frontier of nuclear structure calculations. We mention, for
instance, the recently accomplished exact diagonalziation for double
beta-decay studies in $^{76}$Ge and $^{76}$Se \cite{Cau96,Cau99b}, and
the determination of the $T=1$ two-body matrix elements by fit
\cite{Lis04}.

\begin{figure}[hbpt]
\includegraphics[scale=0.36]{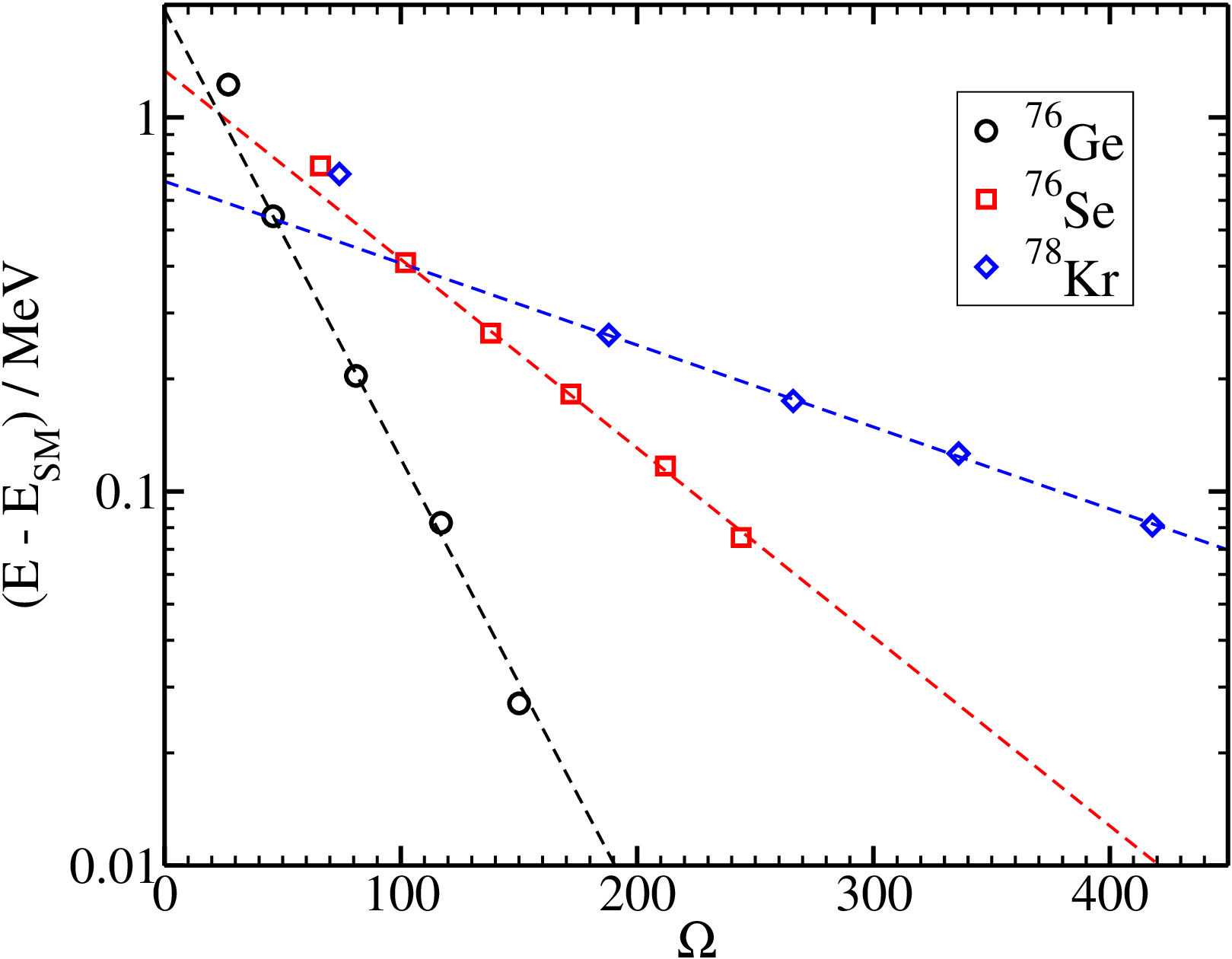}
\caption{\label{fig2}Data points: Ground state energies as a function
of the number of retained states. Dashed lines: exponential fit to data.}
\end{figure}

In this article, we focus on the nuclei $^{76}$Ge,
$^{76}$Se, and $^{78}$Kr, and compute their lowest-lying
states using the $m$-scheme factorization. For Se and Kr, we also use
parity to reduce the dimensionality of the shell-model problem.
Figure~\ref{fig2} shows the energy difference $E(\Omega)-E_{SM}$ of
the ground states as a function of the number $\Omega$ of kept states.
The shell-model energies $E_{SM}$ are determined by an exponential fit
to the numerical data points and are listed in Table~\ref{tab1}.  Note
that the wave function factorization uses a few hundred states out of
up to several hundred thousand proton and neutron states.  Deformed
nuclei typically require more states as deformation is driven by
proton-neutron interactions \cite{Fed79}, and the corresponding wave
functions exhibit stronger entanglement between proton states and
neutron states. Nevertheless, the wave function factorization is very
efficient. For $^{78}$Kr, for instance, the model space contains about
320,000 Slater determinants for the protons and neutrons, and
these combine to about a $1.4\times 10^9$ dimensional configuration
space of definite parity. The largest eigenvalue problem we solve
within the factorization uses only about 400 correlated states for
protons and neutrons, has a much smaller dimension of $3.5\times
10^6$, and the Hamiltonian matrix contains about $2.6\times 10^9$
nonzero matrix elements.  These numbers are small compared to what
exact diagonalizations for comparably large shell problems would
require \cite{Cau02}.  Note also that the angular momentum expectation
value $\langle \hat{J}^2\rangle$ is practically an integer value for
the largest numbers $\Omega$ of retained states.

\begin{table}[b]
\caption{\label{tab1}Shell-model energy $E_{SM}$ of the ground state
and the excitation energies $E_{2^+}^{\rm th}$ of the lowest
$J^\pi=2^+$ state compared to experimental results $E_{2^+}^{\rm
exp}$. Unit of energy is MeV.}
\begin{indented}
\lineup
\item[]\begin{tabular}{@{}*{4}{l}}
\br                              
Nucleus&\m$E_{SM}$&$E_{2^+}^{\rm th}$&$E_{2^+}^{\rm exp}$\cr 
\mr
$^{76}$Ge&-57.23  &0.70& 0.563\cr
$^{76}$Se&-75.74  &0.46& 0.559\cr
$^{78}$Kr&-96.06  &0.44& 0.455\cr
\br
\end{tabular}
\end{indented}
\end{table}

We compute excited states as a by-product of the ground state
factorization.  This approach yields the optimal proton (neutron)
configurations of the excited states in the presence of the neutron
(proton) configuration of the ground state. The excitation energies of
the first excited $J^\pi=2^+$ states are also listed in
Table~\ref{tab1}. They are obtained from the largest calculations we
performed and are expected to be upper bounds (as the ground state
converges more rapidly than the excited states), and the theoretical
uncertainty is about 150 keV. The comparison with the experimental
results is reasonable, and for higher precision of the theoretical
results one would need to target excited states directly \cite{Pap04}.

\section{Summary}

In this article we described recent progress with the DMRG and the
wave function factorization. Within the DMRG, we obtained fairly 
well-converged results for large scale nuclear structure problems using
realistic interactions. Keeping 100-200 states yielded energy
deviations of a few hundred keV. The improved convergence rests mainly
on the implementation of the finite DMRG algorithm, an improved
ordering of single-particle orbitals, and the use of well-selected
states at the beginning of the finite algorithm. Further progress
should be possible by optimizing these aspects. 

We used the wave function factorization for nuclear structure studies
of a few $0f_{5/2}\,1p\,0g_{9/2}$ shell nuclei. Typically, one needs
only a few hundred correlated proton and neutron states for an
accurate approximation of low-lying states, and the error involved is
of the order of 100 keV. Our results show that accurate structure
calculations for $fpg$-shell nuclei and tests of the effective
interactions are possible at a fraction of the effort associated with 
exact diagonalizations.

\ack 

We thank J. Dukelsky and S. Pittel for valuable discussions and
suggestions, and F. Nowacki for providing us with the $fpg$-shell
interaction. Part of this work was performed during TP's stay at the
Institute for Nuclear Theory, University of Washington, during the
program INT-04-3.  This research was supported in part by the
U.S. Department of Energy under Contract Nos.\ DE-FG02-96ER40963
(University of Tennessee) and DE-AC05-00OR22725 with UT-Battelle, LLC
(Oak Ridge National Laboratory (ORNL)). We acknowledge the use of
resources at the Center for Computational Sciences at ORNL.


\section*{References}

\begin{thebibliography}{99}
%
\bibitem{Nav00a}
P. Navr{\'a}til, J. P. Vary, and B. R. Barrett,
\prl {\bf 84}, 5728 (2000), nucl-th/0004058.
%
\bibitem{Nav02}
P. Navr{\'a}til and W. E. Ormand,
\prl {\bf 88}, 152502 (2002).
%
\bibitem{Cau94}
E. Caurier, A. P. Zuker, A. Poves, and G. Mart{\'\i}nez-Pinedo,
\prc {\bf 50}, 225 (1994), nucl-th/9307001.
%
\bibitem{Cau99}
E. Caurier, G. Mart{\'\i}nez-Pinedo, F. Nowacki, A. Poves, J. Retamosa, and
A. P. Zuker,
\prc {\bf 59}, 2033 (1999), nucl-th/9809068.
%
\bibitem{Hon02}
M. Honma, T. Otsuka, B. A. Brown, and T. Mizusaki,
\prc {\bf 65}, 061301(R) (2002), nucl-th/0205033.
%
\bibitem{Hor94}
M. Horoi, B. A. Brown, and V. Zelevinsky,
\prc {\bf 50}, R2274 (1994), nucl-th/9406004.
%
\bibitem{And01}
F. Andreozzi and A. Porrino,
\JPG {\bf 27}, 845 (2001).
%
\bibitem{Gue02}
V. G. Gueorguiev, W. E. Ormand, C. W. Johnson, and J. P. Draayer,
\prc {\bf 65}, 024314 (2002), nucl-th/0110047.
%
\bibitem{Hon95}
M. Honma, T. Mizusaki, and T. Otsuka,
\prl{\bf 75}, 1284 (1995).
%
\bibitem{Whi92}
S. R. White,
\prl{\bf 69}, 2863 (1992).
%
\bibitem{Whi93}
S. R. White,
\prb {\bf 48}, 10345 (1993).
%
\bibitem{Pap03}
T. Papenbrock and D. J. Dean,
\prc {\bf 67}, 051303(R) (2003), nucl-th/0301006.
%
\bibitem{Pes99}
I. Peschel, X. Wang, M. Kaulke, and K. Hallberg (Eds.),
{\it Density-Matrix Renormalization Group} (Springer-Verlag, Berlin 1999).
%
\bibitem{Duk04}
J. Dukelsky and S. Pittel,
\RPP {\bf 67}, 513 (2004), cond-mat/0404212.
%
\bibitem{Duk02}
J. Dukelsky, S. Pittel, S. S. Dimitrova, and M. V. Stoitsov,
\prc {\bf 65}, 054319 (2002), nucl-th/0202048.
%
\bibitem{Dim02}
S. S. Dimitrova, S. Pittel, J. Dukelsky, and M. V. Stoitsov,
nucl-th/0207025.
%
\bibitem{Leg03}
{\"O}. Legeza and J. S{\'o}lyom,
\prb {\bf 68}, 195116 (2003),
cond-mat/0305336.
%
\bibitem{Xia96}
T. Xiang, 
\prb {\bf 53}, 10445 (1996),
cond-mat/9603020.
%
\bibitem{Bro88}
B. A. Brown and B. H. Wildenthal,
{\it Ann. Rev. Nucl. Part. Sci.} {\bf 38}, 29 (1988).
%
\bibitem{Kuo68}
T. T. S. Kuo and G. E. Brown,
\NP {\bf A114}, 241 (1968).
%
\bibitem{Pov80}
A. Poves and A. P. Zuker,
{\it Phys. Rep.} {\bf 70}, 235 (1980).
%
\bibitem{Pap04}
T. Papenbrock, A. Juodagalvis, and D. J. Dean,
\prc {\bf 69}, 024312 (2004), nucl-th/0308027.
%
\bibitem{Shu83}
J. Shurpin, T. T. S. Kuo, and D. Strottman,
\NP {\bf A408}, 310 (1983).
%
\bibitem{Cau96}
E. Caurier, F. Nowacki, A. Poves, and J. Retamosa,
\prl {\bf 77}, 1954 (1996), nucl-th/9601017.
%
\bibitem{Cau99b}
E. Caurier, F. Nowacki, A. Poves, and J. Retamosa,
\NP  {\bf A654}, 973c (1999).
%
\bibitem{Lis04}
 A. F. Lisetskiy, B. A. Brown, M. Horoi, and H. Grawe,
\prc {\bf 70}, 044314 (2004), nucl-th/0402082.
%
\bibitem{Fed79}
P. Federman and S. Pittel
\prc {\bf 20}, 820 (1979).
%
\bibitem{Cau02}
E. Caurier and G. Mart{\'\i}nez-Pinedo,
\NP  {\bf A704}, 60c (2002).
%

\end{thebibliography}
\end{document}